\documentclass{article}
\usepackage[utf8]{inputenc}
\usepackage{commath}
\usepackage{mathptmx,graphicx}
\usepackage[utf8]{inputenc}
\usepackage{bm}% bold math
\usepackage{braket}
\usepackage[T1]{fontenc} 
\usepackage{float}
\usepackage{amsmath,amsfonts,amssymb,eqnarray}  
\usepackage{color, soul}
\usepackage{graphicx}  
\usepackage{dblfloatfix} 
\usepackage{array,booktabs}
\usepackage[pdftex,hypertexnames=false]{hyperref} 
\usepackage[affil-it]{authblk}
\usepackage[font={small}]{caption}
\usepackage{lineno}
\usepackage{soul}
\usepackage{physics}
\usepackage{braket}

\title{Efficient room-temperature molecular single-photon sources for quantum key distribution}

\author{Ghulam Murtaza$^{1,2,3}$}
\author{Maja Colautti$^{2,3,4}$}
\author{Michael Hilke$^{5,6}$}
\author{Pietro Lombardi$^{2,4}$}
\author{Francesco Saverio Cataliotti$^{2,4,6}$}
\author{Alessandro Zavatta$^{2,4,7}$}
\author{Davide Bacco$^{7,8}$}
\author{Costanza Toninelli$^{2,4,*}$}
\affil{$^1$ \small Physics Department - University of Naples, via Cinthia 21, Fuorigrotta 80126, Italy}
\affil{$^2$ \small National Institute of Optics (CNR-INO), Via Nello Carrara 1, Sesto F.no 50019, Italy}
\affil{$^3$\small Co-first authors with equal contribution}
\affil{$^4$\small European Laboratory for Non-Linear Spectroscopy (LENS), Via Nello Carrara 1, Sesto F.no 50019, Italy}
\affil{$^5$\small Department of Physics, McGill University, Montréal, QC, Canada, H3A 2T8}
\affil{$^6$\small Department of Physics, University of Florence, 50019 Sesto Fiorentino, Italy}
\affil{$^7$\small QTI SRL, Largo Enrico Fermi, 6 - 50125 Firenze, Italy}
\affil{$^8$\small Department of Photonics Engineering,
Technical University of Denmark, Kgs. Lyngby, Denmark}
\affil{$^*$\small toninelli@lens.unifi.it}

\date{\today}

\begin{document}

\maketitle

\begin{abstract}
%\LaTeX{} Single photons are key to transmitting quantum information over large distances. A paramount application of quantum communication is Quantum Key Distribution (QKD). QKD protocols allow the distribution of cryptographic keys between two or more users, in an information-theoretic secure way, exploiting the laws of quantum physics. Current QKD systems are mainly based on attenuated laser pulses, which limit the overall range of applicability and leave space for eavesdropping possibilities. On the contrary, true and deterministic single-photon source could give a concrete advantage in terms of secret key rate and security. Here we introduce and demonstrate a proof-of-concept QKD system exploiting a molecule-based single photon source operating at room temperature. Leveraging the intrinsic stability, the ultra-low probability of multi-photon events and the high brightness of this two-level system, we can estimate a secret key rate of 40 kbps at 10 dB of channel losses. Our solution paves the way for room-temperature single photon sources for quantum communication protocols. \textcolor{red}{(we need to reduce to 100 words)}\\

Quantum Key Distribution (QKD) allows the distribution of cryptographic keys between multiple users in an information-theoretic secure way, exploiting quantum physics. While current QKD systems are mainly based on attenuated laser pulses, deterministic single-photon sources could give concrete advantages in terms of secret key rate (SKR) and security owing to the negligible
%ultre-low
%\textcolor{blue}{(lower o possiamo azzardare negligible?)}
probability of multi-photon events. Here, we introduce and demonstrate a proof-of-concept QKD system exploiting a molecule-based single-photon source operating at room temperature and emitting at $785\,$nm. With an estimated SKR of 0.5 Mbps, our solution paves the way for room-temperature single-photon sources for quantum communication protocols.
\end{abstract}

%%%%%%%%%%%%%%%%%%%%%%%%%%  body  %%%%%%%%%%%%%%%%%%%%%%%%%%
\section{Introduction}
Quantum key distribution (QKD), i.e. the distribution of cryptographic keys exploiting the principle of quantum mechanics, is the most advanced technology in the field of quantum communication and represents the first concrete step towards the realization of the quantum internet ~\cite{Preskill_1999, Wehner_2018,pirandola2019advances}. In particular, QKD allows two or more users to exchange cryptographic keys which are used to ensure secure data communication in an information-theoretic-secure way.
Although a few companies and startups are already active in the worldwide market, there are multiple factors that are currently limiting a full deployment of QKD technology.
These can be cast in four different categories: the maximum link distance~\cite{chen2020sending,boaron2018secure,yin2016measurement}, the amount of key generation~\cite{islam2017provably,Bacco2019boosting}, the coexistence of quantum signals with classical communication channels~\cite{Bacco2019boosting,wang2020long} and the security parameters of the implemented QKD systems~\cite{pirandola2019advances}.

In fact, current QKD systems (considering only discrete variable systems) are mainly based on so-called weak coherent pulses (WCP), where a coherent quantum state is prepared with a strongly attenuated laser approaching %\textcolor{purple}{(se si mette per pulse al posto di approaching va messo well below)?} NO
the single-photon
%\textcolor{blue}{(per pulse?)} NO
regime to emulate quantized light emission. Since multi-photon events in coherent states are still possible, it is necessary to decrease the amount of information which can leak to the eavesdropper. A possible solution was put forward by using decoy states (i.e., a random change of the laser intensity over time)\cite{lo2005decoy}. Despite being well established in the community, the decoy state method brings a relevant risk through the opening of back doors in the quantum communication systems~\cite{tang2013source}. In contrast, high purity single-photon sources (SPS) bring advantages both in terms of the secret key rate (SKR) and in terms of security. This is due to the very low probability of multi-photon events. 

Few experiments have shown the ability to generate quantum keys by exploiting deterministic single-photon source. Examples include quantum dots\cite{Takemoto2015quantum, Kupko2020,chaiwongkhot2020enhancing,vajner2021quantum,Arakawa2020progress} and color centers in diamond \cite{Leifgen_2014evaluation}, where promising results were obtained in terms of key generation rates as compared to standard QKD systems. However, most of these systems need cryogenic temperatures, which involve higher costs and limited portability. 

Here, we characterize a single-molecule as a triggered SPS operating at room temperature for QKD experiments. In particular, we demonstrate the feasibility of a BB84 protocol with polarization encoding %\textcolor{blue}{which exploits the single photon states generated by the molecule, \st{of
of the single-photon states generated by the molecule, in a free-space laboratory link. We report better or competitive results with respect to the state-of-the-art-experiments at cryogenic or room temperature \cite{Takemoto2015quantum, Kupko2020,chaiwongkhot2020enhancing,zeng2022integrated,Leifgen_2014evaluation}, both in terms of source efficiency and of expected SKR. Furthermore, we present a detailed calculation of the achievable SKR upon optimization of the optical setup, of the nano-photonic configuration and of the integrated molecular emitter, demonstrating a competitive solution even against protocols using decoy states. Considering the expected SKR of $0.5\,$Mbps resulting from the reported measurements and the potential for future improvement, this technology could boost the deployment of single-photon sources for QKD applications and more generally for different quantum communication protocols.

\section{Single-photon source and Experimental Setup}
Single molecules of polyaromatic hydrocarbons (PAH) in suitable
host matrices have proven to be excellent quantum light sources \cite{Wang2019turning,Rezai2018,Basche1992,Orrit1990}. They show bright single-photon emission of high purity even at room temperature \cite{lombardi2020molecule,Pazzagli_2018,Lounis2000, Toninelli2021} and excellent photostability in long-term measurements \cite{Toninelli2010a}. At cryogenic temperature, PAH molecules emit highly indistinguishable photons \cite{Rezai2018, Lombardi2021}, enabling more complex quantum communication protocols \cite{Llewellyn2020chip} or linear optic processing\cite{knill01}.
Molecular quantum emitters are suitable for the integration in hybrid photonic structures too, allowing for an almost $100\%$ collection efficiency \cite{Lee2011,Chu2017}. Furthermore, their great advantage in quantum photonics arises from the simple recipes used for the device preparation, producing large ensembles of nominally identical molecules at low cost.

\begin{figure}[h!]
\centering\includegraphics[width=12cm]{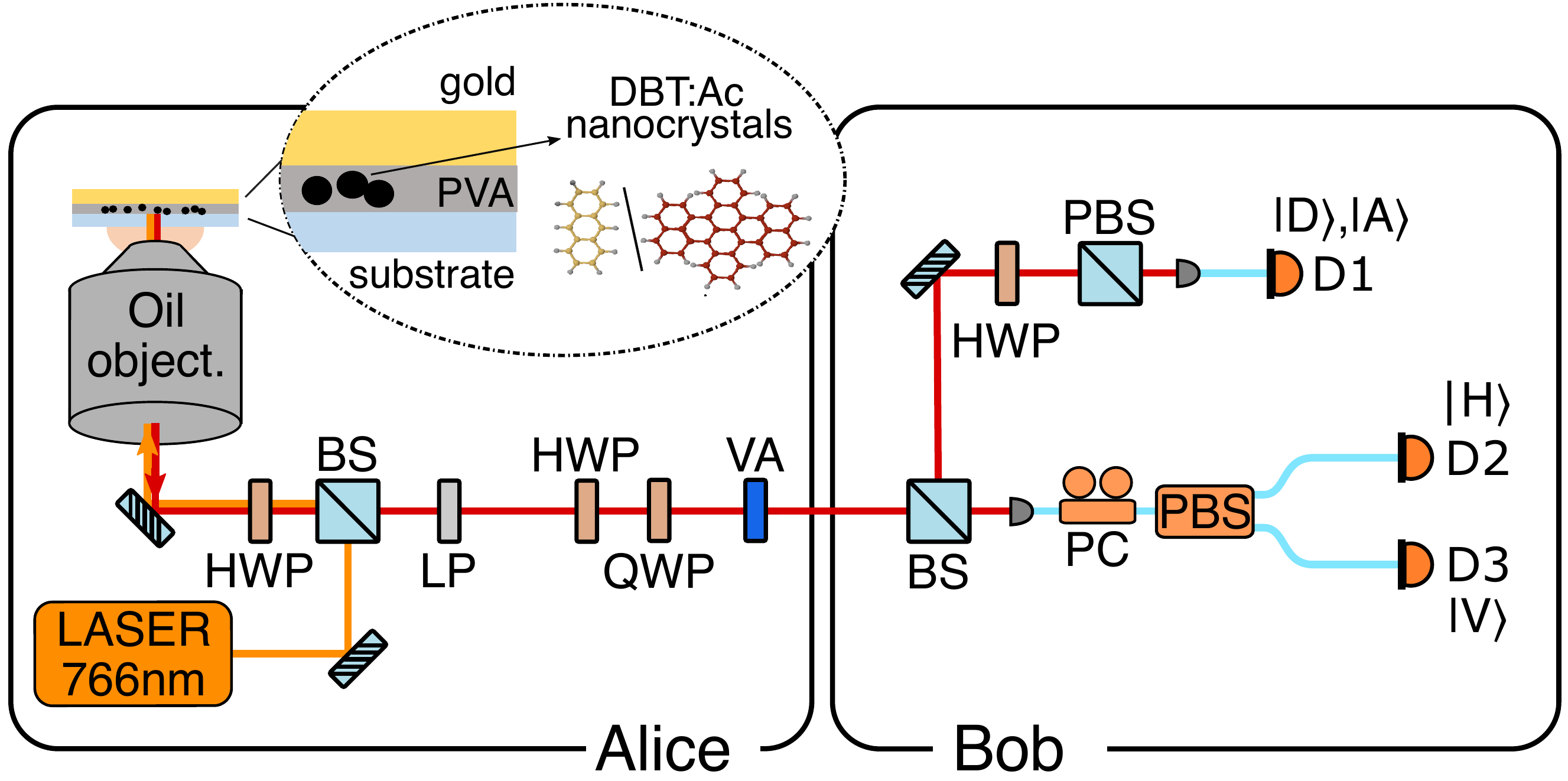}
\caption{\textbf{Experimental testbed for room temperature QKD:} \textbf{Alice)} Single dibenzoterrylene molecules embedded in anthracene nanocrystals (DBT:Ac) are integrated in a planar multi-layer photonic structure for collection enhancement \textbf{(zoom-in)}. An epifluorescence confocal microscope with oil-immersion objective is used to trigger and collect the emission of single-photon packets with 100-picosecond long laser pulses at $766\,$nm and $80\,$MHz repetition rate. The pump laser is filtered-out with a long-pass filter (LP), four quantum states are encoded in the single-photon polarization  ($\ket{H}$, $\ket{V}$, $\ket{D}$ and $\ket{A}$) by means of a half-wave (HWP) and a quarter-wave plate (QWP). Variable attenuators (VA) emulate channel losses. \textbf{Bob)} At the reflection of a beam splitter (BS), a HWP and a polarizing beam splitter (PBS) are used to discriminate between $\ket{D}$ and $\ket{A}$ at the free SPAD D1. At the BS-transmission port, photons are fiber-coupled and a fiber polarization control (PC) is optimized to route the $\ket{H}$ and $\ket{V}$ states at the two outputs of a fiber PBS, respectively connected to fiber SPADS D2 and D3.}\label{fig1}
\end{figure}

In our experiment, single dibenzoterrylene molecules are embedded in anthracene nanocrystals (DBT:Ac) and used as room-temperature SPS (see ref. \cite{Pazzagli2018} for details on the nanocrystals growth). In the zoom-in of Fig.\ref{fig1} the molecular structure of the DBT:Ac system and the sample configuration is shown. The nanocrystals are dispersed on a silica substrate, covered with a thin film of polyvinyl alcohol (PVA) for protection against matrix sublimation, and finally coated with gold. This photonic scheme, in combination with an oil-immersion objective on the substrate side, allows for an enhanced collection efficiency in a robust and planar geometry with a simple fabrication process \cite{Checcucci2017}, which is limited to thin-film deposition and metal coating and result in an overall sample preparation time of about one hour.

The home-made and compact optical setup developed for this work is shown in Fig.\ref{fig1}, where the source, the transmitter (Alice) and the receiver (Bob) parts are highlighted. An electron-multiplying CCD (EMCCD) camera is used to initially image the sample and select the optimal DBT:Ac nanocrystals in terms of fluorescence intensity. Single DBT molecules are then excited confocally through a vibrational state of the electronic transition with a linearly polarized pulsed laser, operating at a central wavelength of $766$-nm and with a $80$-MHz repetition rate. The laser polarization is optimized to match the molecular dipole, which is typically parallel to the substrate plane \cite{Toninelli2010a}, and fluorescence in the range $780-830\,$nm is collected and separated from the pump light via a combination of a $30:70$ (R:T) beam-splitter and a long-pass filter. In this experiment, the excitation laser is also employed directly as a WCP source for comparison with the decoy state method.

Alice station makes use of an achromatic half-wave plate and a quarter-wave plate for the preparation of the quantum state, encoded in the polarization of each single-photon pulse. For this proof-of-concept demonstration the waveplates are manually rotated to switch among the horizontal (H), vertical (V), diagonal (D) and anti-diagonal (A) polarizations, in order to realize the four-state BB84 QKD protocol. The polarization-encoded photons are then sent through a set of neutral density filter attenuators to emulate the effect of channel losses over a broad dB range, hence mimicking the transmission over long distances.

The photons received by Bob are analyzed with a passive choice of the measurement basis. A beam splitter (BS) is used to reflect
%\textcolor{blue}{around 7\% of the \st{some} [maybe add that we set small angle to have almost polarization independent reflectivity?]} NO
photons into a free-space channel, where the combination of a half-wave plate, a polarizing beam-splitter and a free-space Single-Photon Avalance Detector (SPAD, D1) is used for the discrimination between $\ket{D}$ and $\ket{A}$ states. In particular, a manual rotation of the half-wave plate is applied to switch the detection between the two states. The transmission of the BS is coupled to a single-mode fiber
%\textcolor{blue}{instead, where a fiber polarization controller (PC) is optimized to route the $\ket{H}$ and $\ket{V}$ states at the two outputs of a fiber polarizing beam-splitter (PBS), respectively connected to fiber SPADS D2 and D3 \st{ - i like old version better:
polarizing beam-splitter (PBS) where fiber polarization controls (PC) are optimized to discriminate state $\ket{H}$ and $\ket{V}$ at one or the other fiber SPAD connected at the outputs, respectively (D2 and D3). Such fiber setup is also employed in a Hanbury-Brown-and-Twiss (HBT) configuration for the measurement of the second-order correlation function, during the preliminary SPS characterization. Finally, the detected photon counts and arrival times in the four channels are recorded via a multi-port time-tagging system.

%We note in passing that the manual state preparation used in this experiment is preliminary and functional to the following demonstration. The current setup can also be readily upgraded for real-state preparation, just by integrating a fiber-coupled phase modulator for fast and random polarization state transmission \textcolor{blue}{(preparation?)}.

\section{Results}
\begin{figure}[h!]
\centering\includegraphics[width=12.cm]{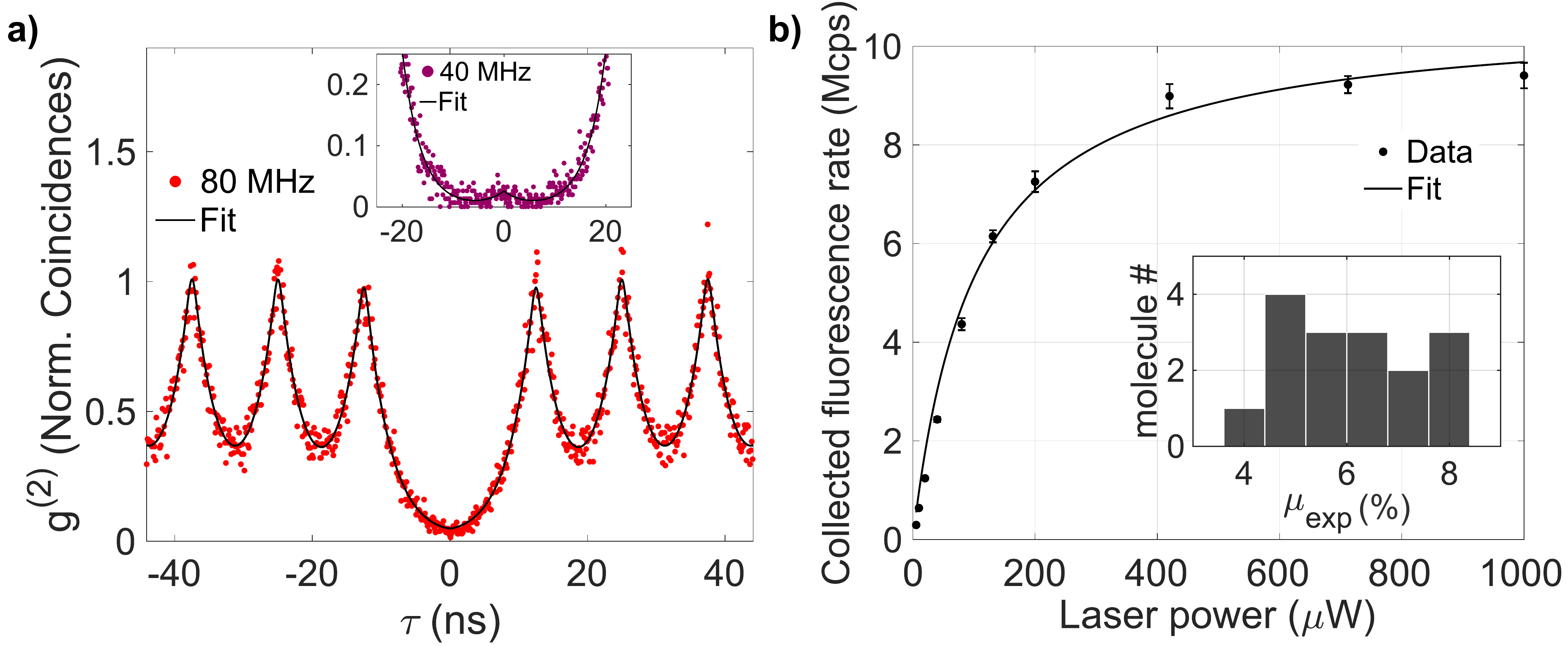}
\caption{\textbf{Characterization of single-photon emission:} \textbf{a)} Normalized histogram of photon coincidences in HBT configuration (red dots) and fit to the data (solid black line) for the second-order correlation function $g^{(2)}(\tau)$. \textbf{inset)} Zoom-in of the central suppressed peak for data measured under $40\,$MHz repetition rate (purple dots) and associated fit (solid black line).  \textbf{b)} Collected single-photon rate as a function of laser pump
power (black dots) and fitted saturation function (black solid line). Distribution of mean photon number values for 16 molecules in different nanocrystals (\textbf{inset}). }\label{fig2}
\end{figure}

As a first step, we characterize the DBT single-photon source in terms of multi-photon probability and collected photon rate, by analyzing the second-order %auto-
correlation function $g^{(2)}(\tau)$ and the collected photon flux as a function of pump power. This is done for different molecules. %These measurements are key to evaluate the purity of the single-photon emission, the emitter off-time due to the long-lived triplet state which directly affects the source efficiency, and the maximum source emission rate
Typical results are shown in Fig.\ref{fig2}. In panel (a), the normalized histogram of the inter-photon arrival times, which approximates $g^{(2)}(\tau)$ for small time delays, is reported for the photon streams collected at D2 and D3 in the HBT configuration. The suppressed peak at zero-time delay gives evidence of the extremely low multi-photon emission probability, and considering the expression\\ $g^{(2)}(\tau)=g^{(2)}(0)exp(-|\tau|/\tau_c)+\sum_nexp(-|\tau+nT|/\tau_c)$, where the $n$-index runs on the order number of the lateral peaks, $\tau_c$ is the dip characteristic time and $T$ the laser repetition period, we retrieve from the best fit to the data $\tau_c=3.6\pm0.1$.
For a more accurate evaluation of the single-photon emission purity we fit the data measured for the same molecule at half of the laser repetition rate (see inset of Fig.\ref{fig2}a) where the suppressed peak can be clearly distinguished, and we obtain  $g^{(2)}(0)=0.02\pm0.01$. 
Considering only Bob's side in Fig.\ref{fig1}, the overall collected single-photon rate is measured at D1 by
summing the contributions from states $\ket{D}$ and $\ket{A}$
and accounting for the BS reflection and to the losses of Bob free space channel, namely the optics $\eta_{opt}\sim 80\%$ and detector efficiency $\eta_{det}=(30\pm2)\%$. Owing to the broad molecule's emission spectrum at room temperature ($\sim50\,$nm), $\eta_{det}$ is experimentally estimated by calibration against power-meter measurements for different laser wavelengths and calculating the weighted average, based on the spectral-intensity distribution. Bob's efficiency is then given by the product $\eta_{\mbox{Bob}}=\eta_{opt}\cdot\eta_{det}$.
%saturation $S=(0.10\pm0.03)mW$
A typical result for the collected photon rate as a function of laser power is shown in Fig.\ref{fig2}(b). The experimental data follow a characteristic saturation behaviour and fit to $R=a+bp+R_\infty p/(p+p_S)$, with $a$ background offset, $b$ laser leakage, $p$ laser power and $p_s$ saturation power, yielding a maximum collected count rate of $R_{\infty}=(10\pm2)\,$Mcp. In particular, the operational pump power for the QKD experiments discussed below corresponds to a saturation parameter $s=P/p_S\sim2$, where $P$ stand for the selected power. This is chosen so as to optimize the trade-off between single-photon purity, quantum bit error rate and source efficiency. 
The resulting mean photon number is obtained by dividing the corresponding collection rate by the laser repetition rate, which for the case of the saturation curve in Fig.\ref{fig2}(b) yields $\mu_{mol}=0.08\pm0.01$ and is among the best values reported for solid-state single-photon sources for QKD \cite{Takemoto2015quantum,Leifgen_2014evaluation,chaiwongkhot2020enhancing}. Repeating the procedure on 16 molecules in different nanocrystals leads to the distribution displayed in the inset of Fig\ref{fig2}(b). The inherent variability is likely due to different  factors, such as the different local crystalline environment at the molecular dipole position (i.e. different distance to the nanocrystal surface and interface effects), as well as the distance to the gold film (see inset of Fig.\ref{fig1}), which provides optimal enhancement for a value of $\sim100\,$nm (see also the Discussion section).

\begin{figure}[h!]
\centering\includegraphics[width=12cm]{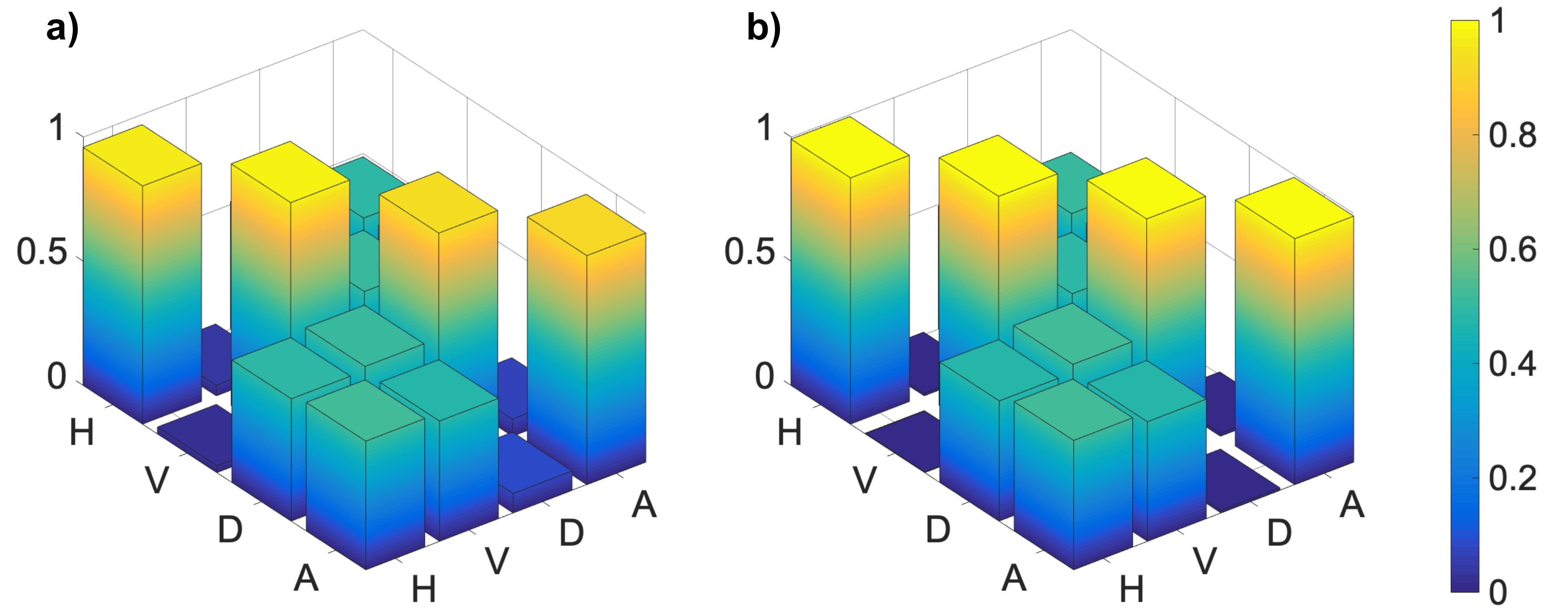}
\caption{\textbf{Outcome distribution} in the output discrimination channels for the set of four input states $\ket{H}$, $\ket{V}$, $\ket{D}$ and $\ket{A}$, in the case of zero channel losses \textbf{(a)} for the single-photon emission - fidelity $F=(99\pm1)\%$ \textbf{(b)} and for the WCP with $\mu=0.5$ - fidelity $F=(99.78\pm0.03)\%$, respectively.}\label{fig3}
\end{figure}

We employ here the previously characterized single-photon pulses as polarization-encoded qubits in a four-state BB84 QKD protocol for the key generation\cite{pirandola2019advances}. For each of the four discrimination channels described in the previous section, the quantum bit error rate (QBER) is evaluated after state preparation of $\ket{H}$, $\ket{V}$, $\ket{D}$ and $\ket{A}$, yielding an average QBER of $3.4\pm 0.2 \%$ in the back-to-back configuration. %$QBER_H=(3.8\pm0.2)\%$, $QBER_V=(2.8\pm 0.2)\%$, $QBER_D=(6.8\pm 0.1)\%$ and $QBER_A=(7.2\pm0.4)\%$, respectively. Note, that the QBERs in the D and A channels are larger because only $3.4\%$ of the transmitted signal is directed to the A and D branch, the rest is left for the H and V branch. 
In Fig.\ref{fig3}(a), the matrix of the normalized counts (using single-photon emission) in each output channel for a given input channel (equivalent to the outcome distribution for the four set of states) is presented in a 3D colour map for the best case scenario of zero channel losses, corresponding to having no attenuator in Fig.\ref{fig1}. In panel (b), we report the state preparation matrix using an attenuated laser (i.e., weak coherent pulses) for  a  mean photon number per pulse of $\mu=0.50\pm0.03$. To quantify the fidelity of the states and the transmission effects we resort to the expression of fidelity $F(p,r)=\langle\sum_i(p_ir_i)^{1/2}\rangle$\cite{cozzolino2019orbital}, where $p_i$ and $r_i$ are the experimental and theoretical elements of the probability distribution for each polarization state, $\langle \cdot \rangle$ stands for the average over the four considered states. We note here, that $p_i$ is obtained from the matrix shown in Fig.\ref{fig3} (experimental outcome distribution) upon normalization (probability distribution). Hence, the fidelity yields, respectively, $(99\pm1)\%$ and $(99.78\pm0.03)\%$ for the single-photon state and the WCP with $\mu=0.5$. This result attests to the robustness of the molecule-based proposed testbed.

\begin{figure}[h!]
\centering\includegraphics[width=12.cm]{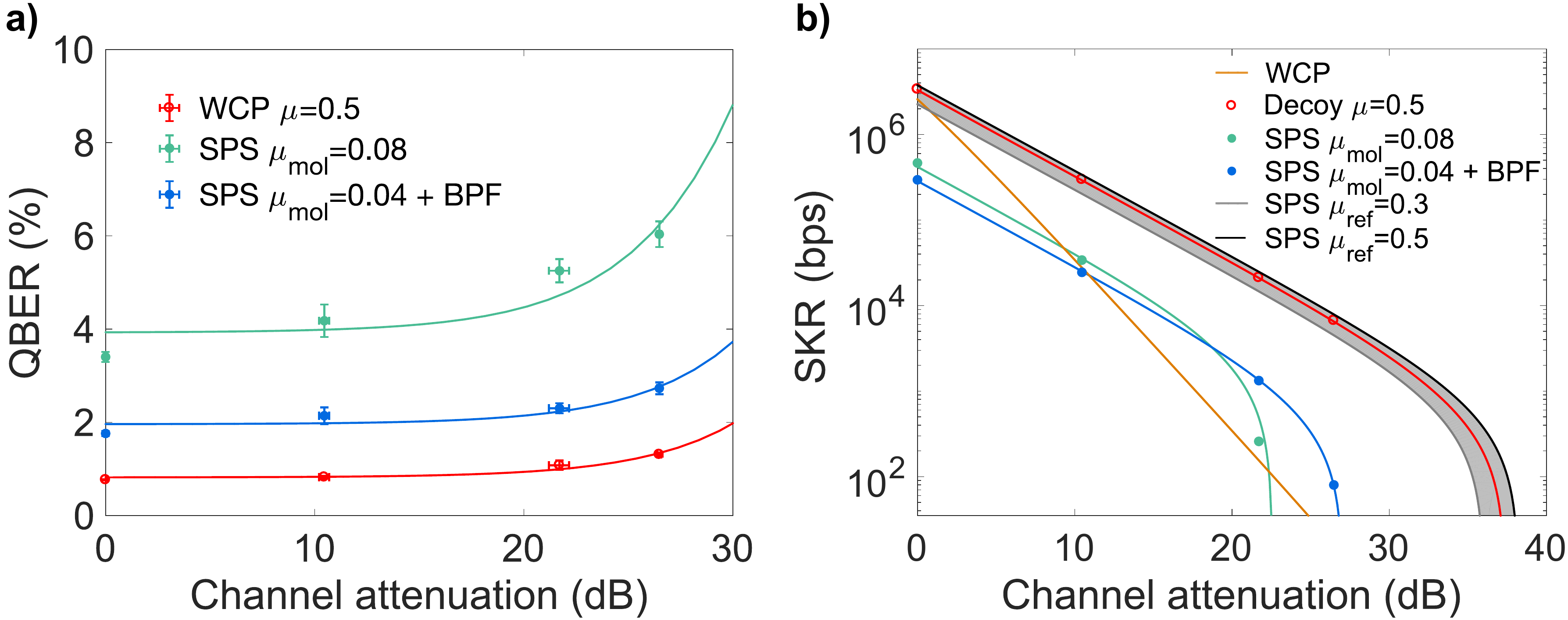}
\caption{\textbf{QKD channel characterization:} \textbf{a)} QBER as a function of total losses for two different molecule-based single-photon sources (with and without band pass filter) and for a WCP source. Scattered points stand for the experimental data while the corresponding lines are fits to equations \eqref{QBER1} and \eqref{QBER2}. \textbf{b)} The lines show the extracted SKRs based on the experimentally determined detection error probability ($e_{det}$) and dark counts following equ. \eqref{SKR_SPS} for the molecule source and the corresponding expressions for the WCP SKR, without decoy and with decoy \cite{ma2005practical}. The scattered points are based on the measured QBERs for different channel attenuations. In addition, we show with the gray and black lines and shaded area the expected SPS-source SKR with an ideal $\mu_{ref}=0.3-0.5$ (assuming the same QBER as in the left figure).}\label{fig4}
\end{figure}

To determine the expected SKR as a function of the channel loss we can experimentally evaluate the corresponding QBERs by inserting  a varying set of attenuators (see Fig.\ref{fig1}). The  weighted average QBER is shown in Fig.\ref{fig4}(a) as a function of the channel loss, $\eta_{\mbox{channel}}$. We show the experimental results for a WCP source at $\mu=0.5$ and two different molecules exhibiting different $\mu_{mol}$ (see inset of Fig.\ref{fig2}b). By using the total loss $\eta=\eta_{\mbox{Bob}}\cdot\eta_{\mbox{channel}}$, we can fit the QBER of the molecule-based source using 
\begin{equation}
  \mbox{QBER}=\frac{P_D/2+e_{det}\,\eta\,\mu}{P_D+\eta\,\mu_{mol}}
  \label{QBER1}
\end{equation} 
to extract a value for the detector dark counts ($P_D$) and the detection error probability ($e_{det}$). The best fits give a $P_D$ in the range $0.4-4\times 10^{-6}$ counts per pulse, while $e_{det}=(3.9\pm0.5)\%$ for a molecule with $\mu_{mol}=0.08\pm 0.01$ counts per pulse.  This can be improved by adding a band pass filter ($40\,$nm), which yields $e_{det}=(2.0\pm0.2)\%$ but to the detriment of $\mu_{mol}=0.04$ counts per pulse. Similarly, for the WCP we can fit the QBER using a slight modification that takes into account the Poisson distribution of the photon number per pulse in the WCP case\cite{ma2005practical}
\begin{equation}
  \mbox{QBER}=\frac{P_D/2+e_{det}(1-e^{-\eta\,\mu})}{P_D+1-e^{-\eta\,\mu}}.
  \label{QBER2}
\end{equation} 
The best fit leads to a similar $P_D$ as above and to $e_{det}=(0.8\pm 0.1)\%$ for $\mu=0.5$ counts per pulse. We are now in the position to evaluate the expected SKR as a function of channel loss. For the SPS case with molecules, multi-photon events are strongly suppressed, as characterized by the second order correlation at zero time delay $g^{(2)}(0)\simeq 0.02$ discussed in more detail above. This leads to a small multi-photon probability given by $P_m=\mu_{mol}^2g^{(2)}(0)/2$ \cite{Waks2002}. To evaluate the SKR in this case we can use for instance the expression in Ref. \cite{Kupko2020}, which reads
\begin{equation}
SKR_{\mbox{SPS}}=\frac{1}{2}P_{click}\left[\beta\,\tau(\mbox{QBER})-f(\mbox{QBER})\,H(\mbox{QBER})\right],
\label{SKR_SPS}
\end{equation}
where the factor $1/2$ stems from the sifting ratio for symmetric basis encoding, $P_{click}=\mu_{mol}\,\eta+P_D$ is the probability of having a click in the detector,  $\beta=(P_{click}-P_m)/P_{click}$ is a correction factor for possible multi-photon events, $H$ is the standard binary Shannon information function, $\tau(x)=-\log_2(1/2+2x-2x^2)$ and $f(QBER)$) is the error correction efficiency \cite{Waks2002}. The $\mbox{SKR}_{\mbox{SPS}}$ together with the experimental data points obtained with single molecule sources at room temperature are shown in Fig.\ref{fig4}(b). 

As a comparison, we also show the SKR that would be obtained using a WCP laser. We compare the efficient vacuum and weak decoy state method proposed by Ma and co-workers (equation (42) in ref. \cite{ma2005practical}), assuming an optimal choice of $\mu \simeq 0.5$ and decoy $\nu \simeq 0.05$. To include our experimental data points, we equate the one photon error rate $e_1=\frac{P_D/2+e_{det}\,\eta}{P_D+\eta}$ and one photon gain $Q_1=(P_D+\eta)\mu e^{-\mu}$ in equations (8) and (9) from \cite{ma2005practical} and use our experimental values for QBER, $\eta$, $e_{det}$, $P_D$, $\mu$ and $\nu$. We also show in the same figure the case when no decoy state is used for the attenuated laser source. This situation is more relevant to the comparison with the SPS, which does not require decoy states to be secure. To compute the simulation of the weak-coherent QKD protocol without decoy, we have used the expression from ref. \cite{scarani2009security}.

For the purpose of illustrating the potential of our room-temperature molecular SPS platform, the expected SKR for an average number of photons per pulse between $\mu_{ref}=0.3-0.5$ is depicted in Fig.\ref{fig4}(b) as an ideal case scenario (assuming the same QBER as in the left figure).  

\section{Discussion}
%%%% Discussion on skr
We observe that the developed single-photon source is already competitive with respect to the attenuated laser case.
%\textcolor{blue}{\st{, especially for equivalently long-distance propagation}}. 
Moreover, the current experimental results appear very promising when compared to other state-of-the-art experiments employing solid-state quantum emitters \cite{Takemoto2015quantum, chaiwongkhot2020enhancing,Kupko2020,Leifgen_2014evaluation}. The proposed molecular source can be advantageous especially when considering its room-temperature operating condition, as opposed to semiconductor epitaxial quantum dots that are used at cryogenic temperatures. %\textcolor{blue}{[true for epitaxial, what about colloidal?]}.

A protocol implementing decoy states with attenuated lasers offers higher SRKs compared to our $8\%$-efficiency source. In Fig.\ref{fig4}(b), the almost one order of magnitude difference between the SKR achieved with the SPS versus the WCP with decoy state, clearly shows that the use of our SPS platform for QKD applications would benefit from further optimization. The picture changes quickly if one considers higher efficiency values, albeit with the same $g^{(2)}(0)$. The SKR extrapolated for the ideal case scenarios of molecules exhibiting a $\mu_{ref}$ between 0.3 and 0.5 demonstrates that molecule-based SPSs could bring a key advantage upon optimization, as discussed in more detail below. 
%For example, in Chaiwongkhot et al. (ref. \cite{chaiwongkhot2020enhancing}), the authors employ a quantum dot embedded in a nanowire to demonstrate a single-photon SKR comparable or even better than the result for the decoy state case, despite the analogous values for the laser repetition rate, mean photon number (both for the single-photon and decoy state) and purity with our experiment.
If we focus on the single-photon case, we already achieved an expected SKR of $\sim0.5\,$Mbps for the back-to-back configuration (violet circles) and, as a second reference point, $\sim80\,$bps for $27\,$dB channel losses (light-blue circles). These values are already better or competitive with the best ones obtained in the literature for cryogenic SPSs \cite{Takemoto2015quantum, chaiwongkhot2020enhancing,Kupko2020} or room temperature SPSs \cite{Leifgen_2014evaluation,zeng2022integrated}. In some of these implementations, longer telecom wavelengths are used and include quantum dots, nanowire quantum dots, colour centers in diamond and epitaxial quantum dots. Longer wavelengths lead to larger losses on Bob's side due to less efficient photon counting at these wavelengths \cite{Takemoto2015quantum}. They are instead optimal for fiber-based communication networks.

\begin{figure}[h!]
\centering\includegraphics[width=12.cm]{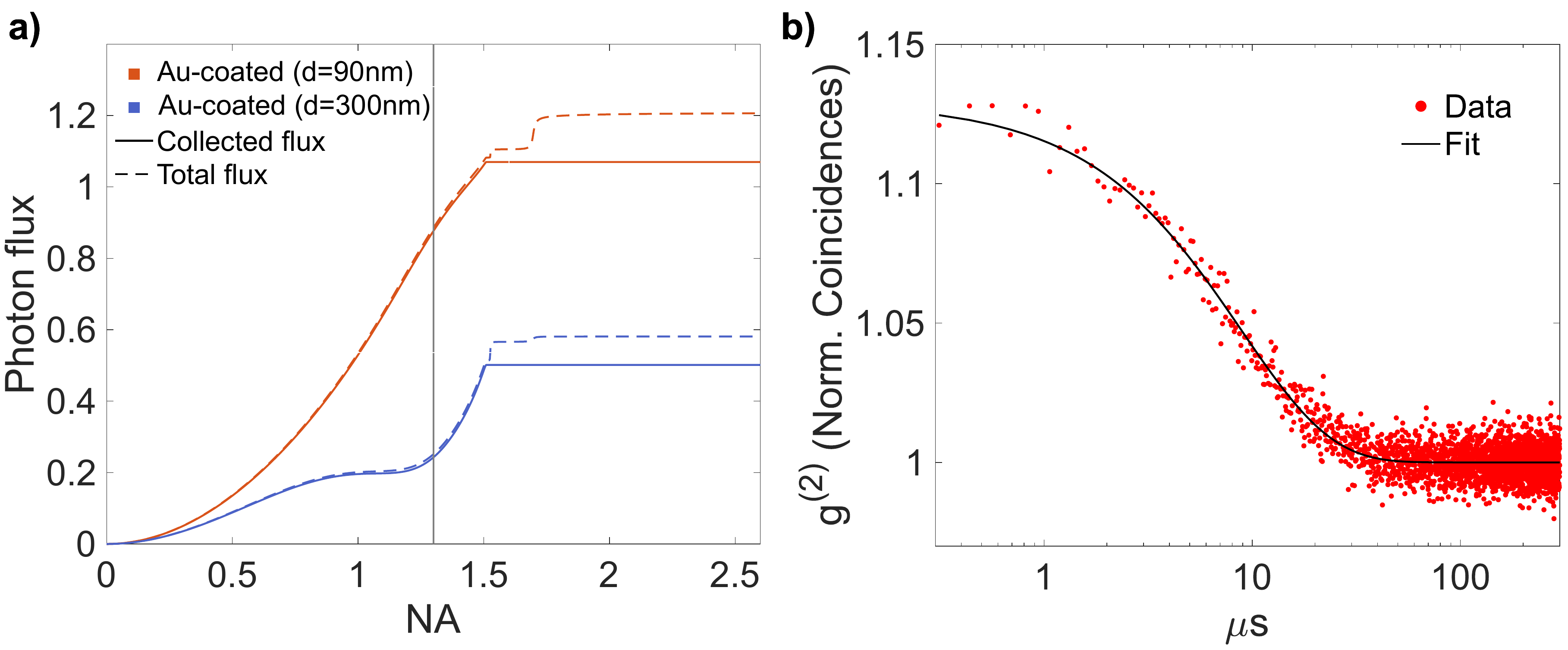}
\caption{\textbf{a)} Normalized photon flux as a function of the objective numerical aperture NA. The photon flux is calculated for the sample multilayer shown in Fig\ref{fig1} and composed of $120\,$nm of gold film and $500\,$nm-thick nanocrystals in PVA on a glass substrate, and is then normalized to the emission in a homogeneus medium of anthracene. The collected flux (solid line) is compared to the total flux (dashed line) integrated in the full $4\pi$ solid angle around the dipole emitter. Colours stand for different dipole distances $d$ from the gold layer. \textbf{b)} Experimental data (dots) for the $g^{(2)}$ at long times and fit (line) describing a three-level system dynamics \cite{Nicolet2007a}. }\label{fig5}
\end{figure}

As discussed above, to achieve an even higher SKR with a SPS the mean photon number has to be increased, as illustrated by  $\mu_{ref}$ in Fig.\ref{fig4}(b). The molecular mean photon numbers can be enhanced upon realistic optimization of the experimental configuration.
%This case represents for the molecule-based single-photon emission an optimal mean-photon number of $\mu_{ref}=0.5$, and shows hypothetical results which are actually comparable with the decoy state protocol. In order to understand the assumptions supporting $\mu_{ref}$ as realistic reference value,
First, we need to consider the different contributions to $\mu_{mol}=\eta_{opt}\eta_{col}\eta_{mol}$, which are the efficiency of the optics on Alice side ($\eta_{opt}$ - see Fig.\ref{fig1}), of the collection ($\eta_{col}$) and of the molecule emission ($\eta_{mol}$), respectively. In particular, $\eta_{opt}=0.54\pm 0.02$ is given by the measured transmittivity of all the components along the optical path from the sample to the attenuators (Alice side). The evaluation of $\eta_{col}$ is a geometrical factor based on the simulation of the angular emission profile and is calculated numerically by modelling the sample multilayer presented in the inset of Fig.\ref{fig1}. The molecule emission dipole is placed into a nanocrystal with thickness of $500\,$nm. In Fig.\ref{fig5}(a), the resulting collected flux is compared to the total flux including also non-radiative losses for two values of the dipole distance from the gold layer, i.e. the optimal condition $d_1$ for enhanced collection, and a reasonably assumed worst case $d_2$.  Correspondingly, we can extrapolate the two bounds for the collection efficiency for our objective numerical aperture (NA=1.3 - grey vertical line in the figure) yielding $\eta_{col,1}=0.74\pm0.06$ and $\eta_{col,2}=0.44\pm0.08$. 
As final contribution to $\mu_{mol}$ to consider, $\eta_{mol}=QY\eta_{pump}ON_{\%}$ depends on the quantum yield of the emitter $QY$, on the pumping efficiency $\eta_{pump}$ and on the ON-times of the molecule $ON_{\%}$, defined as percentage of emission events over excitation cycles. This latter parameter  can be evaluated from the $g^{(2)}(\tau)$ at long times, which is shown in panel (b) and is measured under CW excitation for clarity. In particular, from the drop of normalized counts at long times we can extrapolate the average trapping time in the dark triplet state and hence  $ON_{\%}=0.77\pm0.05$ \cite{Bernard1993a,Nicolet2007a}.
%In particular, the drop of normalized counts at tens of millisecond delay time is the signature of the transition to a long-lived triplet state of the molecule which is responsible for OFF-times in the emission. Using the fit function $g^{(2)}(\tau)=1+Ae^{-\sigma\tau}$, and considering... \textcolor{red}{(either we write all the relations or just put a reference - maybe better since its too many details for a qkd paper)}, we find $ON_{\%}=80\%$. 
Secondly, considering that $\eta_{pump}=P_{e,\infty}\frac{R(P)}{R_{\infty}}$, we assume as excitation probability $P_{e,\infty}=0.75$ which is the maximum value at room temperature \cite{Schofield2018}, and from the ratio between the collected rate at the operational saturation parameter $R(P)$ and the maximum rate $R_{\infty}$, we obtain $\eta_{pump}=0.47\pm0.07$. Hence, we can extrapolate the quantum efficiency for the two considered cases of dipole distance, yielding $QY_{ex,1}=0.6\pm0.1$ and $QY_{ex,1}=0.9\pm^{0.1}_{0.3}$. These values are
%is similar to the result obtained for DBT molecules in spincoated Ac \cite{kwadrin2012}, but much 
lower than the almost unitary $QY$ displayed by several PAH molecules \cite{Buchler2005}, but this is motivated by the room-temperature operation at which the $QY$ can be strongly reduced owing to temperature dependent non-radiative decay pathways. \\
%However, it is also well known that solid-state emitters as colour centers in diamond \cite{mohtashami2013} and organic molecules \cite{Chu2017} typically exhibit inhomogeneously distributed quantum efficiencies which can result in very diverse emission intensities. We can hence reasonably assume that optimal values for DBT can realistically amount to $QY^*_{DBT}=2\times QY_{DBT,ex}=0.8$.
Finally, based on the estimation of all the involved experimental parameters contributing to $\mu_{mol}$, we can extrapolate the reference value $\mu_{ref}$ in the ideal case scenario in terms of sample configuration and setup optimization. If we consider the demonstrated $99\%$ collection efficiency for organic molecules in Ref.\cite{Chu2017}, a realistic improvement of the optics efficiency up to $\eta^*_{opt}=90\%$, and the upper bound to $\eta^*_{pump}=P_{e,\infty}$, we can obtain $\mu_{ref}=\eta^*_{opt}\eta^*_{col}\eta^*_{pump}QY_{exp,i}ON_{\%}$, yielding $\mu_{ref,1}=0.31\pm0.06$ and $\mu_{ref,2}=0.5\pm0.2$ for the two estimated values of the $QY_{exp,i}$.

This considered, the combination of molecule-based emitters and an optimal optical configuration would bring beyond the break-even point and become advantageous with respect to the use of weak coherent pulses and decoy states.

\section{Conclusion}
In conclusion, we successfully implemented a proof-of-concept QKD setup employing a deterministic single-photon source operating at room temperature. The results, in terms of expected SKR ($0.5\,$Mbps at zero losses), are competitive with state-of-the-art experiments - at cryogenic and room temperature - and can be further improved in the near future by optimizing the nano-photonics of the sample configuration and the optical setup. In this regard, taking into account all the experimental contributions to the overall source efficiency and analyzing in detail the margin for improvement, we have evaluated the achievable SKR of the molecular emitter demonstrating the potential advantages in using the generated single-photon states even compared to the decoy state performances.\\
Thanks to the room-temperature operation and long-term photostability of the emitter, the proposed hybrid technology is especially interesting for satellite quantum communication. An ultra-compact and cost-effective QKD setup configuration can be envisioned for the integration in a next-coming satellite quantum-encrypted network, or in a CubeSat for preliminary testing and experiments. Moreover, upon down-conversion to telecom wavelengths \cite{dalio2022pure,zaske2012visible} the system can also be efficiently operated in fiber communication networks. Considering a future upgrade to real-time state-preparation and -measurement, this platform paves the ground towards a practical use of truly single photons for QKD applications, both in terrestrial and spatial links, and for quantum communication protocols in general.

\subsubsection*{Acknowledgments}
D.B. and C.T. conceived the research, M.C. and P.L. designed the experiment.
G.M. performed the measurements; M.H and M.C performed the data analysis, M.C, M.H. and D.B wrote the manuscript with critical feedback from all authors.

\subsubsection*{Funding}
This project has received funding from the EraNET Cofund Initiatives QuantERA within the European Union's Horizon 2020 research and innovation program grant agreement No. 731473 (project ORQUID), from the FET-OPEN-RIA grant STORMYTUNE (Grant Agree- ment No. 899587) of the European Commission, and from the EMPIR programme (project 17FUN06, SIQUST and 20FUN05 SEQUME), co-financed by the Participating States and from the European Union’s Horizon 2020 research and innovation programme.

\subsubsection*{Disclosures}
\noindent The authors declare no conflicts of interest.

\subsubsection*{Data Availability Statement}
The data that support the findings of this study are available
from the corresponding author upon reasonable request.

%%%%%%%%%% If using BibTeX:
%\bibliography{biblio}{}
\bibliographystyle{ieeetr}

%\printbibliography

\end{document}